\documentclass[a4paper,11pt]{article}

\usepackage{adjustbox}
\usepackage[utf8]{inputenc}
\usepackage{amsmath}

\usepackage{adjustbox}

\begin{document}
\title{Dark Matter as Screened ordinary Matter}
\author{C. D. Froggatt, Glasgow University\\
H.B.Nielsen, Niels Bohr Institute}

\maketitle
{\bf Proceedings for 
the 28th International workshop ‘’What
comes beyond the standard models’’, 2025, July 7 - 16.}


\begin{abstract}
  We look at our since long studied model for dark matter as being
  pearls of a speculated new vacuum containing highly compressed
  ordinary matter, 
  with so much ordinary in it that the content of ordinary matter
  in the dark matter pearls dominate. Most dark matter models have the dark
  matter consisting mainly of new-physics-matter such as WIMPs being
  supersymmetric partners of possibly known particles or, as in Maxim Khlopov's
  model, a doubly negatively charged new-physics-particle with a
  helium nucleus attached. But usually the new-physics-matter makes
  up weightwise the major content. It is only in our model that the ordinary matter content in the dark matter dominates. We here expose some weak
  phenomenological evidence that, in truth, dark matter should be
  of the type with a dominant component of ordinary matter (weightwise),
  thus favouring as the typical example our previously so much studied
  vacuum type 2 model. The main such evidence is that we manage a fit to data
  in which the 3.5 keV X-rays, presumed to result from dark matter,
  come {\em both} from collisions of dark matter with dark matter and
  from dark matter with ordinary matter! Both mechanisms are of so similar an
  order of magnitude that they are both seen, indicating that their similarity
  is due to a significant similarity between dark with ordinary matter. The fact that the amounts of ordinary and dark matter
  {\em only} deviate by a factor 6 points in the same direction.
  Using the information obtained from this fitting, we develop our speculation that the main content weightwise of dark matter is ordinary matter to the very DAMA experiment. Actually we found three spots on the sky in which we fit the observed production of 3.5 keV X-rays with ordinary + dark
  scattering.  
\end{abstract}

\section{Introduction}

Most theoretical models on the market for dark matter involve new physics
in one form or the other, because seemingly the Standard Model alone
cannot explain or even provide an appropriate possibility for a model
for the dark matter. However, if the story about the several
phases would be right, {\em we} would only need new physics in explaining that the parameters of the Standard Model were fine tuned to ensure that more than
one phase of the vacuum would be competitive and present in Nature. 
In most models of dark matter, even according to weight, the
major constituents of the dark matter  
are made up from new physics - i.e. speculated new particles such as is the
case in typical WIMP models in which the new physics particles could be
susy-partners of e.g. gauge bosons. Even in Maxim Khlopov's model in which
the main ingredient is a doubly negatively charged particle, this new physics
doubly negatively charged particle is expected to be so heavy that its mass
dominates over the accompanying ordinary matter helium nucleus. The models
that use axion like particles ALP's are of course also having dark matter
dominated by new physics in as far as no axion has as yet been found. But here
of course division as to what the dark matter is made from after weight
is a bit more delicate, in as far as the genuine constituents are supposed to be
extremely light and the ``coldness'' of the dark matter is only supposed
to come in by bose-statistics effects. But, in any case, it is mostly new
physics making up the dark matter.

As far as we know it is only in our own model \cite{Dark1,Dark2,Tunguska,
  supernova,Corfu2017,Corfu2019,theline,Bled20,Bled21,extension,Bled22,
  Corfu2022,Corfu2022A,Bled23,Corfu23} that the dark matter is
made up after weight dominantly from {\em ordinary matter}. We
have to admit that the utterly important ingredient in our model, that
the dark matter consists of bubbles of a new physics vacuum, is of course new physics. However we should keep in mind that if the computer analysis
of what is called the Columbia plot should end up telling that there is
a phase transition in QCD not hitherto taken so seriously, it could
become old physics that we use.

The main attitude of the present article is to consider a slight generalization
of our model in the sense of prescribing a class of models.
In these models the dark matter is dominantly ordinary
matter, only made ``dark'' in some way, such as e.g. having the nuclei very
strongly screened so that they no longer interact with charged particles
as strongly as usual atoms. In our model there is actually such a strong
screening, because we make the assumption that the vacuum inside the
pearls has a lower potential for nucleons than the outside vacuum.
This then namely means that a lot of nuclei are pulled into the interior
and pressed together by the tension in the domain wall separating the two
different vacua - the latter is supposed to have a tension of the order
of the third power of a few MeV - and the nuclei pull the electrons with them electrically. Thus we get a high electron density in the pearl and therefore
a very strong screening. It is really this screening, which makes the nuclei
interact so weakly with electrical objects or atoms, that ensures the dark
matter in our model really can be arranged to interact so little that we must
call it ``dark''.

Our model of a dark matter particle or pearl is a cluster of screened
nuclei kept inside a skin or domaine wall.

\includegraphics{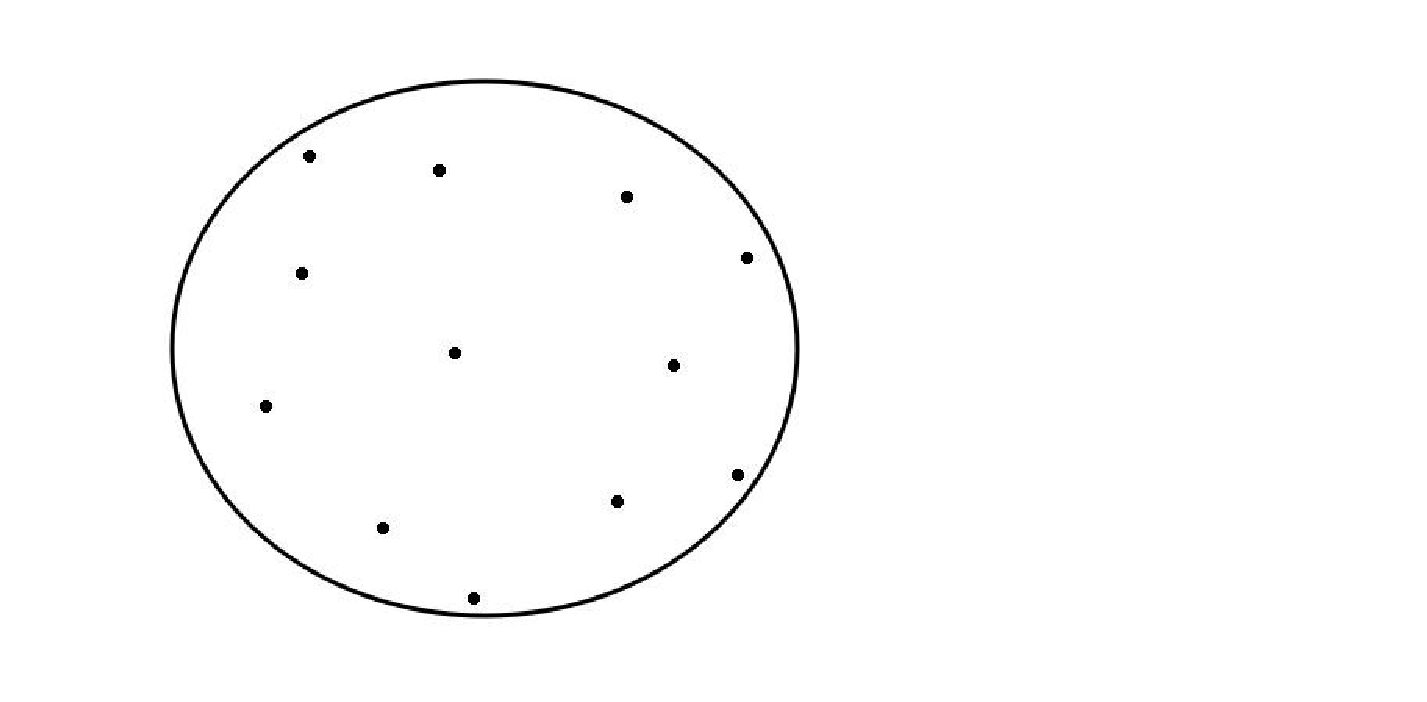}

\section{Interference factor}

Let us list and compare some numbers, which we can collect concerning the
non-gravitational observations of dark matter, and seek to put them in the
form of a ratio  $\frac{``\sigma''}{m}$ of an effective cross section with
significant scattering angle $``\sigma''$ divided by the mass of the particle
having this cross section $m$.

For a given density $\rho$ of a medium in which a particle penetrates,
the stopping length is at least crudely given by
\begin{eqnarray}
  \hbox{``stopping length''} &\approx &\frac{1}{\frac{``\sigma''}{m}
     *\rho}
\end{eqnarray}
and so, for a fixed medium, we can consider this ratio $\frac{``\sigma''}{m}$
a measure for stopping power. We take as our starting point the approximation
that the dark matter pearl consists effectively just of a swarm of strongly
electrically screened nuclei, which separately scatter independently -
so we shall first consider the interference between hitting different
constituent nuclei
in the next paragraph. In this approximation we shall have this ratio being
the same for clusters of the constituents as for the constituent nucleus itself
as long, of course, that the $m$ in the denominator is the mass of the cluster.

But now the crucial point is that for elastic scattering of a dark matter pearl
there will occur interference  between all the cases of a constituent from one
of the two colliding pearls meeting a constituent from the other one. Also if an ordinary
matter nucleus hits a dark matter pearl we have for the elastic
scattering interference between the hitting of all the constituents.

Thus we have interference corrections to this ratio $\frac{``\sigma''}{m}$:
\begin{eqnarray}
  DM+DM \rightarrow DM+DM, \hbox{ correction }
  \frac{``\sigma''}{m}&\rightarrow&
  \frac{``\sigma''}{m}*\hbox{``\# nuclei in DM pearl''}^2\nonumber\\
   DM+OM \rightarrow DM+OM, \hbox{ correction }
  \frac{``\sigma''}{m}&\rightarrow&
  \frac{``\sigma''}{m}*\hbox{``\# nuclei in DM pearl''}\nonumber
\end{eqnarray}

\subsubsection{How to think on our quantities}

For the following it may be good to have in mind that given the quantity
$\frac{\sigma}{m}$, the stopping power is crudely invariant under splitting
up of the particles in the medium into essentially non-interacting
constituents, as for example the splitting
\begin{eqnarray}
  DM &\hbox{split into }& scnu + scnu +...+ scnu
  \end{eqnarray}
where $scnu$ stands for screened nuclei.

For seeing this one has to have in mind that a hit particle typically
gets a velocity of the order of the velocity of the penetrating particle and
thus the momentum loss becomes of the order of $m_{hit}v$.
Of course non-relativistically the number $n$ of constituents, $scnu$'s,
will equal
\begin{eqnarray}
  n &=& \frac{m_{DM}}{m_{scnu}}.
  \end{eqnarray}
where $scnu$ stands for screened nuclei.


So
\begin{align*}
  \hbox{whether } \frac{\sigma}{m} &\; \hbox{taken to } &
  \frac{\sigma_{DM +DM}}{m_{DM}} & \; or & \frac{\sigma_{DM+scnu}}{m_{DM}}\\
      \hbox{``stopping f.''}&\;\propto&\frac{\sigma_{DM+DM}}{m_{DM}}*
      \rho_{\# \;DM}*vm_{DM} & \; or & \frac{\sigma_{DM+scnu}}{m_{DM}}*
      \rho_{\# \; scnu}*vm_{scnu}\\
      \hbox{I.e. } \hbox{``stopping f.''} & \propto&
 \frac{\sigma_{DM+DM}}{m_{DM}}\rho_{mass}v&\; or&
      \frac{\sigma_{DM+scnu}}{m_{DM}}*
      \rho_{mass}*v,\\
      \hbox{I.e. } \hbox{``stopping f.''} & \propto&
 \frac{\sigma}{m_{DM}}\rho_{mass}v&\; or&
      \frac{\sigma}{m_{DM}}*
      \rho_{mass}*v
\end{align*}
I.e. you get the same stopping force ``stopping f.'' in a model wherein
the DM are genuine particles and $\sigma = \sigma_{DM+DM}$ as in our model in
which the dark matter pearls are ideally loosely bound constituents, called
$scnu$, provided you put $\sigma = \sigma_{DM + scnu}$ and provided you
ignore quantum mechanics, meaning the interference between the scattering on
different constituents.

Also the construction of the ratio $\frac{\sigma}{m} $ is invariant under the
splitting of the particle into its constituents, say $scnu$'s, by going to the
ideally loosely bound cluster, in the sense that
\begin{eqnarray}
  \frac{\sigma_{DM + something}}{m_{DM}}&=&\frac{\sigma_{scnu+something}}{m_{scnu}}
  \hbox{(classically)}.
  \end{eqnarray}

The above results were only true {\bf classically}, but let us postpone the quantum discussion till we have settled how to interpret the Correa data.

\subsection{Correa}

The self-interaction of the dark matter, as extracted from the dwarf galaxy
studies of Camilla Correa \cite{Correa}, are already in the form we go for,
in as far as
we believe she found
\begin{eqnarray}
  \frac{``\sigma''}{m}_{Correa}&=&\frac{``\sigma''_{DM+DM}}{m_{DM}}\\
  &\approx& \frac{K}{v^2}\\
  \hbox{where }K &=& 10^{10}m^4/s^2/kg\\
  \hbox{With $1m=5.07 *10^{15}GeV^{-1}$}&,& \hbox{$1kg=5.625 *10^{26}GeV$ }:
  \nonumber\\
  K &=& 5.08*10^{-13}GeV^{-3}/(m^4/s^2/kg)*10^{10}m^4/s^2/kg\nonumber\\
  &=& 5.08*10^{-3}GeV^{-3}\\
  \hbox{With } v=3.2*10^5m/s \hbox{ then }\frac{``\sigma''}{m}|_{DM+DM}&=&0.1
  m^2/kg,\label{correasm}\\
  &=& 4.57*10^3GeV^{-3}
\end{eqnarray}


Process
\begin{eqnarray}
  DM+DM &\rightarrow& DM +DM \hbox{(elastic)}
\end{eqnarray}

But now we are in the present article studying the hypothesis, that
the dark matter pearls are ideally loosely bound clusters of some
screened nuclei $scnu$, so that what really happens in the collisions
of dark matter pearls is the collisions of these constituents. The effect
of shadowing is supposed small. Now we have to contemplate what Correa
by her analysis has really measured, we must expect that she has
indeed measured how  fast the dark matter particles in the various
dwarf galaxies are being stopped by the dark matter density present.
That means that, provided our model of loose bound states is right, we can
interpret her values for the ratio $\frac{\sigma}{m}$ as
\begin{eqnarray}
  \frac{\sigma}{m}|_{Correa} &=& \frac{\sigma_{DM +scnu}}{m_{DM}}\\
  &=& \frac{\sigma_{scnu+scnu}}{m_{scnu}} (\hbox{classically}).
\end{eqnarray}

But now switching on quantum mechanics, so that one has positive interference
between all the $n^2$ possibilities for an $scnu$ in one pearl to
interact with an $scnu$ in the other pearl in the collision, the ratio
f for the Correa quantity $\frac{\sigma}{m}|_{Correa}$  above
gets screwed up by a factor $n^2$ to
\begin{eqnarray}
  \frac{\sigma}{m}|_{Correa}&=& n^2 *\frac{\sigma_{scnu+scnu}}{m_{scnu}}
  \hbox{(quantum mechanically)}.
  \end{eqnarray}

\subsection{Cline-Frey}

From the Cline-Frey fit \cite{Cline-Frey} we extracted an average for the
numbers, which we believed
could be fitted with DM+DM scattering, while we left out the items supposedly
rather due to mainly DM+OM scattering (here OM means ordinary matter).
\begin{eqnarray}
  \left (
  \frac{N\sigma_{DM+DM\rightarrow ---+3.5 keV}}{m_{DM}^2}\right )_{exp}&=&
  (1.0 \pm 0.2)*10^{23} cm^2 /kg^2\\
  &=& (1.0\pm 0.2)*10^{19}m^2/kg^2.
\end{eqnarray}

Taking it that the dark matter consists of subparticles - presumably nuclei -
with masses $A m_N$ and kinetic energies thus $A m_N v^2/2$ we may, if all
the energy goes into the 3.5 keV line, produce per collision of the
subparticle get $N$ photons of 3.5 keV, where
\begin{eqnarray}
  N &\approx & A m_n*v^2/(2* 3.5 keV)\\
  \hbox{and } \frac{\sigma}{M}|_{per \; 3.5 keV} &=& \left (
  \frac{N\sigma}{M^2}\right )*3.5 keV/v^2\label{cac}\\
  &=& 10^{19}m^2/kg^2*(3.5 *1.6 *10^{-19}*10^3J/v^2)\\
  &=& 3500Jm^2/kg^2/v^2.
\end{eqnarray}
For $v=3.2*10^5m/s$ we then have  
\begin{eqnarray}
 \frac{\sigma}{M}|_{per \; 3.5 keV}
  &=& 3.5 *10^{-7}m^2/kg.\label{cfsm}
\end{eqnarray}

Process
\begin{eqnarray}
  DM+DM&\rightarrow& DM+DM+ph(3.5 keV)+...
\end{eqnarray}

Notice that in (\ref{cac}) above the ansatz for the mass
$M =Am_n $ dropped out, and so $\frac{\sigma}{M}|_{per \; 3.5 keV}$ gives
the number 3.5 keV photons produced by penetration of 1$kg/m^2$.

Actually we can write the quantity measured by the Cline-Frey analysis
in the following three ways {\bf classically}:
\begin{eqnarray}
   \left (
  \frac{N\sigma_{\rightarrow...+3.5 keV}}{M^2}\right )_{exp}&=&
  (1.0 \pm 0.2)*10^{23} cm^2 /kg^2\\
  &=& (1.0\pm 0.2)*10^{19}m^2/kg^2\\
   &=& \frac{\frac{1/2 *m_{DM}*v^2}{3.5 keV}*\sigma_{DM+DM\rightarrow ...3.5keV}}
       {m_{DM}^2}\\
  &=& \frac{\frac{1/2 *m_{scnu}*v^2}{3.5 keV}*\sigma_{DM+scnu\rightarrow ...3.5keV}}
       {m_{DM}m_{scnu}}\\
       &=& \frac{\frac{1/2 *m_{scnu}*v^2}{3.5 keV}*
         \sigma_{scnu+scnu\rightarrow ...3.5keV}}
       {m_{scnu}^2}
  \end{eqnarray}
For instance the last version simplifies to
\begin{eqnarray}
    \left (
    \frac{N\sigma_{\rightarrow...+3.5 keV}}{M^2}\right )_{exp}&=&
     \frac{\frac{1/2*v^2}{3.5 keV}*
         \sigma_{scnu+scnu\rightarrow ...3.5keV}}
       {m_{scnu}}\\
  \end{eqnarray}

This was still {\bf classically}.

When we switch on quantum mechanics we get positive interference between
scattering on the different constituents $scnu$ in the same dark matter pearl
$DM$, unless the $scnu$ participating got marked in some way so as to make
the interference impossible. We must suppose that, once an excitation of the
electron-system has happened by a hole quasi-electron pair having been
produced, one of the $scnu$'s hitting each other has been marked so that there
remain for these 3.5 keV producing events a positive interference between the
$scnu$ particles in just {\bf one} of the two colliding DM's. Thus only
one factor $n$ (the number of constituents) will occur as the correction of the
classical result to the quantum one:
\begin{eqnarray}
   \left (
   \frac{N\sigma_{\rightarrow...+3.5 keV}}{M^2}\right )_{exp}&=&
   n*\frac{1/2*v^2}{3.5 keV}*
     \frac{\sigma_{scnu+scnu\rightarrow ...3.5keV}}
       {m_{scnu}}
\end{eqnarray}

If $\frac{\sigma_{scnu+scnu\rightarrow ...+3.5 keV}}{m_{scnu}} $ has the
inverse square dependence on the velocity which we like to assume
\begin{eqnarray}
  \frac{\sigma_{scnu+scnu\rightarrow ...+3.5 keV}}{m_{scnu}}&=&
  \frac{K_{scnu+scnu\rightarrow ...+3.5 keV}}{v^2},
  \end{eqnarray}
  then  
  \begin{eqnarray}
  \left (
  \frac{N\sigma_{\rightarrow...+3.5 keV}}{M^2}\right )_{exp}&=& \frac{1}{2 *3.5 keV}
  *n*K_{scnu+scnu\rightarrow ...+3.5 keV}.
  \end{eqnarray}

\subsection{Reaching DAMA}

If we assume that the $\frac{``\sigma''}{M}$ for dark matter on
ordinary matter is just so that a dark matter particle gets essentially stopped just in the depth of DAMA of 1400 m, and an estimated stone density of
$\rho = 3000kg/m^3$, then
\begin{eqnarray}
  \frac{``\sigma ''_{DM+OM}}{M}&\simeq&\frac{1}{1400m * 3000 kg/m^3}\\
  &=& \frac{1}{4.2*10^6kg/m^2}\\
  &=& 2.38 *10^{-7}m^2/kg.\label{damasm}
\end{eqnarray}
Process
\begin{eqnarray}
  DM +OM &\rightarrow& DM+OM \hbox{(mainly elastic)}
\end{eqnarray}

Strictly speaking we should not take it that the stopping length just
makes the dark matter pearls stop at the depth 1400m of DAMA-LIBRA, since that
would be an unlikely coincidence. Rather we should take it that the stopping
length is so much
smaller
than the depth of DAMA that the probability of the
particles stopping just at DAMA - and thus giving a seasonal modulated signal
at DAMA - could just explain the lack of efficiency (our factor $1/2 *10^9$ in  (\ref{ratio}) below
leaves a factor $10^9$ to
be explained by the deviation of the DAMA depth from the depth where the
highest number of pearls stop).

Let us also remark that we shall see in section \ref{ast} below that from the
sign of the seasonal effect observed in DAMA-LIBRA it is needed that
the depth of DAMA is deeper than the dominant stopping depth.

Classically we get the same penetration depth, 
if we think of $m_{OM}$ as the one divided out i.e. $M=M_{om}$, whether we use,
\begin{eqnarray}
	\frac{``\sigma ''_{DM+OM}}{m_{OM}} & or &  \frac{``\sigma ''_{scnu+OM}}{m_{OM}},
\end{eqnarray}
because the higher number of $ scnu$ is compensated for by a lower
momentum loss  by using $m_{scnu}v$ than $m_{DM}v$.
Also oppositely if we think of the divided out $M$ as being put
$m_{DM}$ or $m_{scnu}$ the difference gets divided away classically.
But quantum mechanically
we have an interference between the $n$ constituent $scnu$'s in the same pearl,
and thus
  \begin{eqnarray}
  \frac{``\sigma ''_{DM+OM}}{M}&\simeq&\frac{1}{1400m * 3000 kg/m^3}\\
  &=& \frac{1}{4.2*10^6kg/m^2}\\
  &=& 2.38 *10^{-7}m^2/kg.\label{damasm2}\\
  &=& n\frac{``\sigma ''_{scnu+OM}}{m_{scnu}}. 
\end{eqnarray}

\subsection{The three exceptional places}

In three exceptional places we claim that the 3.5  keV line arises mainly from
dark matter colliding with ordinary matter. We speculated then on physical
grounds in our rather thinly filled dark matter pearls that, counted after
weight, the rate of 3.5 keV line photon production should be the same for
$DM+DM$ as for $DM+OM$. We found support for this assumption of approximate equality
most simply by believing to have found that in the outskirts of the Perseus
Galaxy Cluster, where the ratio of dark to ordinary matter is close to unity,
there is a ``kink'' signalling that the dominant production mechanism for
3.5 keV photons shifts from $DM+DM$ to $DM+OM$. Believing this we assume that
dark matter being hit by nuclei inside another dark matter particle or by
nuclei just present in the ordinary matter would have the same cross
section $\sigma$ per 3.5 keV photon produced. Interpreting the $M$ in the
denominator as the mass of the dark matter particle, that here could either
hit an ordinary or a dark-matter-contained nucleus,  we should be allowed to
use the same $\frac{``\sigma''}{M}$ for the dark matter hitting ordinary matter
as it hitting dark matter. Thus
\begin{eqnarray}
\frac{``\sigma''}{M}&=& 3.5*10^{-7}m^2/kg.\label{excsm}
\end{eqnarray}

Process
\begin{eqnarray}
  DM+OM &\rightarrow& DM+OM +ph(3.5 keV)+...
  \end{eqnarray}

This means we just get the same as we already described under Cline Frey.

Even quantum mechanically we get the same as under Cline-Frey because we
suppose that in the Cline Frey case there could anyway only be interference in
one of the two colliding dark matter pearls, because of the constituent $scnu$
in one of them was marked by having made a 3.5 keV X-ray or more correctly
some hole electron pair.

Thus we have again so to say
\begin{eqnarray}
	\left (
	\frac{N\sigma_{DM +OM\rightarrow...+3.5 keV}}{M^2}\right )_{exp}&=&
	n*\frac{1/2*v^2}{3.5 keV}*
	\frac{\sigma_{OM+scnu\rightarrow ...3.5keV}}
	{m_{scnu}}
\end{eqnarray} 

\subsection{The efficiency of getting 3.5 keV line in DAMA}

We have estimated how big a fraction of the kinetic energy of the incoming
dark matter
would be observed as signals in the DAMA-LIBRA experiments and obtained the
result that it is about $2*10^{-9}$ times the impact kinetic energy of the
dark matter coming in. In the philosophy that all the kinetic energy gets
converted into the 3.5 keV line - which is of course an overestimate -
we would here in the present over-idealized discussion take it that
the only other way to dispense with the kinetic energy is by stopping the dark
matter by elastic scattering. Then we would have to say that the elastic
scattering for $DM + OM$ should be $\frac{1}{2}*10^9$ times bigger than the
inelastic one (by our assumption of only the 3.5 keV line taking the energy
meaning $DM+OM \rightarrow DM+OM +ph(3,5 keV) +...$.).

So we would conclude in terms of the processes we study

\begin{eqnarray} \label{ratio}
  \frac{\frac{``\sigma''}{M}|_{DM+OM(elastic)}}{
    \frac{``\sigma''}{M}|_{DM+OM\; (with \; ph(3.5 keV))}}&=& \frac{1}{2}*10^9,
  \label{effsm}
\end{eqnarray}
if all inelastic energy is converted into 3.5 keV X-rays
But this is not at all likely for the supposed low energies / low velocities
of the dark pearls (see below in section \ref{ast}), because
for low velocities the collisions between the lighter nuclei can hardly
deliver 3.5 keV at velocities such as 300 km/s. For example:
\begin{eqnarray}
  \hbox{For } 300\, km/s &:&\hbox{thresholds }:
  \nonumber\\
  \hbox{proton H}&:& 1/2\, keV\\
  \hbox{helium He}&:& 2\,keV\\
  \hbox{carbon C}&:& 6\, keV\\
  \hbox{sodium Na}&:& 11.5\, keV\\
  \hbox{iodine I}&:&63.5\, keV\\
  \hbox{Thresholds for  }3.5\, keV¥:¥\nonumber\\
  \hbox{proton H} &:& v = 300\, km/s *\sqrt{7}=794\, km/s\\
  \hbox{helium He }&:& v= 300\, km/s *\sqrt{3.5/2}=397\, km/s\\
  \hbox{carbon C }&:& v= 300\, km/s *\sqrt{3.5/6} = 229\, km/s\\
  \hbox{sodium Na } &:& v=300\,km/s*\sqrt{3.5/11.5}= 166 km/s\\
  \hbox{iodine I }&:&v=400km/s *\sqrt{3.5/63.5}=70 km/s
  \end{eqnarray}

If the dark matter pearls mainly consist of light nuclei - below
carbon say - and we notice that for protection against the
cosmic rays the underground experiments have to be of the order
of a km down from the earth surface, then all the experiments are under
the threshold for easily producing 3.5keV. Thus this production is easily
very strongly suppressed even what the non-modulating part of the signal
is concerned. If we take it that the modulating velocity of the Earth of the order
30 km/s is about 10\% of the typical galactic velocity, the energy
at the tails of the dark pearl tracks, which are responsible for the modulation
observation have only about 10\% of the energy of the incoming dark matter
beam and so the typical velocity $300km/s/\sqrt{10} = 95 km/s$.

Ratio of rates for {\bf low velocity dark matter pearls} :
\begin{eqnarray}
  DM+OM &\rightarrow& DM+OM\\
  DM+OM&\rightarrow &DM+OM+ ph(3.5 keV)...
  \end{eqnarray}
and for this we have found the $1/2*10^9$; but for the {\bf high
velocity presumably more relevant for the galactic clusters included by
Cline and Frey this ratio could be of order unity.}

\subsubsection{Comment on the Efficiency}

Let us immediately comment that we shall not take this efficiency as
seen by DAMA as coming from the whole track with the typical
velocity of the dark matter particles being say 300 km/s. In fact we namely show
that the part of the track that matters for the seasonal variation
to be only the tiny tail, where the velocity is very slow such
as to be almost stopping. The point is that we argue 
in section \ref{ast} that in the region of a track in which the velocity is high the energy deposited per unit length 
of the track arising from the stopping is constant independent of the velocity.
Then you cannot see the velocity on the track except at the very end, where a
fast track extends longer than a slow particle track. This means
that the seasonal varying rate observed at DAMA will correspond to very low
velocity dark matter and thus likely to be below the threshold in energy for making the 3.5 keV line. Then production of a 3.5 keV photon would only come by statistical fluctuations and might be estimated by some Boltzmann factor, which could give a very low rate.

\subsection{Discussion}
The main idea to seek some regularity in the above values of  $\frac{``\sigma''}{M}$ is to think of
the dark matter pearl as effectively being a collection of charged but screened
nuclei. The number is really only to be an effective one and probably this
effective number of nuclei is much smaller than the true number of nuclei.
In fact we shall rather think of the nuclei and their associated electrons
forming some electric field and that this electric field around the dark matter
pearl is approximated, say, by an effective nucleus with an appropriate number of
charges. 

We may imagine that the different say nuclei in the effective
collection have their interactions interfere maximally constructively.Then
the total cross section for a process rate will be increased by the number
of effective particles $n$. In fact, if there is no interference
the $n$ constituents will just produce $n$ separate contributions,
but if we have interference the total contribution will be $n^2$ times a single
contribution and thus $n^2/n =n$ times what we have without interference.

Now we shall also have in mind, that if we have up to two dark matter pearls involved in the scattering, then we can have interference involving the constituents of both pearls and thus, if each has $n$ effective constituents, get an interference correction up to even $n^2$,

We expect that, if we have a 3.5 keV photon in the final state, it may have come
from one specific nucleus in one af the dark matter pearls colliding and thus
prevent the presence of the factor $n^2$ from interference in both the dark matter
particles. The observability of the X-ray might make the emitting nucleus
observed and thus prevent the interference in the pearl containing the nucleus. Also of course we can at most get
as many factors $n$ as there are dark matter pearls in the collision.

For the presumably elastic scattering, as observed in Correa's dwarf galaxy
studies, we should expect to have a factor $n^2$ compared to having no
interference. If we believe that the stopping of the dark matter reaching DAMA
just stops after 1400 m, due to elastically dominated DM+OM collisions, we
clearly expect the dominant term to have just one factor $n$ from interference
going on in the only dark matter pearl in the collision(s).

Similarly we expect that the Cline-Frey value for DM+DM inelasticity giving
3.5 keV radiation will contain an interference enhancement factor $n$. This
is  because the
presence of the 3.5 keV quantum has prevented - by it potentially being
associated with pointing out a specific nucleus in the pearl as the one hit -
the interference among the nuclei in one of the two colliding dark
matter pearls.

These three items  or rather the last two, DAMA versus Cline-Frey, can be
considered a success for our hypothesis that the  cross section relative
to mass $\frac{``\sigma''}{M}$ should be the same when exchanging OM and DM.
Also the quantum interference correction, $n$ in this case, does not
distinguish between OM and DM. 
The values of  $\frac{``\sigma''}{M}$ for DAMA and for Cline-Frey
are indeed very close, $(3.5\;  \hbox{and}\; 2.38)*10^{-7}m^2/kg$ for
respectively
Cline Frey and DAMA.

Using the Correa value for  $\frac{``\sigma''}{M}$ we then
get $n \simeq\frac{0.1m^2/kg}{3*10^{-7}m^2/kg}=
3*10^5$ or for the ratio of the $\sigma/m$ for the two supposedly
elastic processes (\ref{correasm}, \ref{damasm}),
\begin{eqnarray}
  \frac{\frac{\sigma}{m}|_{Correa}}{\frac{\sigma}{m}|_{DAMA}}&=& \frac{0.1m^2/kg}
       {2.38*10^{-7}m^2/kg}\\
       &=& 4.2*10^5.
\end{eqnarray}
This ratio should be equal to the number of ``atoms'' in a dark matter pearl,
since we want to explain the difference between the two numbers as due to
(positive) interference between $n$ ``atoms'' in such a pearl.

We should be able to use this ``measurement'' of the number $n$
of constituent nuclei in the pearl to estimate the size of the pearl, at
least if we somehow guess the atomic weight of the constituent nuclei.

By requiring the dark matter pearl to have a homolumo gap of 3.5 keV responsible for the observed X-ray line and thus a Fermi momentum $p_f = 3.3  MeV$ for the electrons inside the pearl, a crude dimensional argument suggests the density of the pearl material is $\rho_B= 5.2 *10^{11}kg/m^3$. Using this value we obtain estimates for the radius  $R$ and the cube root of the surface tension $ S^{1/3}$ of the pearl for two different proposals for the atomic weight $A$ of the constituents:
\begin{eqnarray}
  \hbox{For } A=12 : \; M_{pearl}&=& A*1.66*10^{-27}kg*4.2*10^5\\
  &=& 8.37*10^{-21}kg\\
  \hbox{giving } ``Volume'' &=& \frac{8.37*10^{-21}kg}{5.2*10^{11}kg/m^3}\\
  &=& 1.61*10^{-32}m^3\\
  \hbox{so radius } R &=& \sqrt[3]{1.61*10^{-32}m^3*3/(4\pi)}\\
  &=& \sqrt[3]{3.84*10^{-33}m^3}\\
  &=& 1.57*10^{-11}m;\\
  \hbox{Extrapolating from } R=10^{-10}m &\sim& S^{1/3}=8MeV\\
  S^{1/3}&=&8MeV*\sqrt[3]{\frac{1.57*10^{-11}m}{10^{-10}m}}\\
  &=& 4.3 MeV\\
  *****&***&*****\nonumber\\
   \hbox{For } A=100 : \; M_{pearl}&=& A*1.66*10^{-27}kg*4.2*10^5\\
  &=& 6.97*10^{-20}kg\\
  \hbox{giving } ``Volume'' &=& \frac{6.97*10^{-20}kg}{5.2*10^{11}kg/m^3}\\
  &=& 1.34*10^{-31}m^3\\
  \hbox{so radius } R &=& \sqrt[3]{1.34*10^{-31}m^3*3/(4\pi)}\\
  &=& \sqrt[3]{3.20*10^{-32}m^3}\\
  &=& 3.17*10^{-11}m;\\
  S^{1/3}&=&8MeV *\sqrt[3]{\frac{3.17*10^{-11}m}{10^{-10}m}}\\
  &=&5.45 MeV
  \end{eqnarray}

These values of the cube root of the tension $S^{1/3}$ should be compared with
the value obtained from the straight line fit in the article 
´´Ontological fluctuating Lattice'' \cite{scales1,scales2,scales3, ofi}
in the same issue as this article of the Bled Workshop Proceedings. In fact
since the action for a surface (in Minkowski space a three dimensional track)
in the lattice theory is supposed to be related to the third power
of the link size $a$, i.e. the coefficient to the three dimensional
surface action is proportional to $a^3$, we expect in the straight line
rule for the energy scales, that the energy scale for the surface tension
$S^{1/3}$ shall be three steps, meaning three factors of 220.584, under the
``fermion tip'' scale, which is at $2.06*10^4 GeV$. That is to say we expect
from the straight line model the cubic root of the tension to be:
\begin{eqnarray}
  S^{1/3} &=& \frac{\hbox{``fermion tip''}}{(``step-factor'')^3}\\
  &=& \frac{2.06*10^4 GeV}{220.584^3}\\
  &=& 1.92 MeV.
  \end{eqnarray}

The exact match of this prediction would e.g. be reached with an even lower $A$
than our 12, but one shall consider it a good  agreement taking into account
the crude calculation.  



\subsection{Suggested Solution}

Ignoring at first the efficiency number above, we can understand the
numbers above like this:

There are two different electric fields associated with the dark matter pearls
\begin{itemize}
\item There is a screened Coulomb potential around each nucleus in the pearl.
  Because of the high electron density it is quite short range.
\item There is Coulomb field very similar to that of huge nucleus sitting
  in its atom. It is field much like the one in a rather thin spherical
  condenser. This is caused by the pull of the pearl skin on the nuclei
  inside this skin. They are pressed inward towards the center of the pearl
  by the skin,
  and for balance there has to be electrons outside driving the nuclei
  the opposite way.
  \end{itemize}

It is now the idea that in the collisions observed by Correa in the dwarf
galaxies it is the field from the electrons just outside the skin and the
nuclei just inside the skin that dominates and that the scattering is
mainly elastic. The field is built up from many nuclei etc. and from the
point of view of these nuclei the scattering becomes the result of strong
interference
between  the different nucleus combinations. This strong interference
allows the interaction of just elastic scattering of dark matter on dark matter
to become appreciably larger than processes without this interference.

It can only be for this elastic scattering of two dark matter pearls that you
can expect these fields in the surroundings to work in full interference,
because having only one dark matter pearl when we look for DM + OM can at
least not have the $n^2$ interference and also the interference will be
spoiled by
the excitation of the pearl which would probably carry sign of one or another
region in the pearl having been excited.

In fact we shall suggest that the
types of scatterings listed
above with 3.5 keV photon production or DM+OM scattering 
shall
only have essential interference between
screened nuclei inside {\em one} dark matter pearl,
so that the quantum enhancement will only be one factor $n$, and not
$n^2$ as for the dark + dark elastic scattering (the Correa case). 
So we should in principle be able to calculate
 DM+OM scattering with 3.5 keV production as if we had only
separate screened nuclei instead of dark matter, except that we should
multiply by a single $n$ factor namely only for the interference between
the nuclei inside
just one of the dark matter pearls.

However, there is a threshold effect that may spoil completely this simple
thought when concerning the production of the 3.5 keV line: If the OM nuclei
hitting the dark matter are not sufficiently fast, it will be kinematically
impossible to reach the kinetic energy threshold with collisions with the
velocity $v$ in question. This means that when the velocity is below a
threshold, that for say Na nuclei is about 100 km/s formally, 3.5 keV X-rays
cannot be produced. The effective threshold is probably a bit higher and the
cutting off can be smoothed out. On the other hand iodine would have a bit
higher threshold. 

Let us now look at the three places on the sky, where we think that the
3.5 keV line comes from dark matter hitting ordinary matter and the situation
when the dark matter pearls are about stopping at the DAMA experiment, and
especially note the velocity of the collisions

\begin{itemize}
\item{{\bf PCO}} The Perseus Cluster Outskirts:

  In the Perseus Cluster the temperature of the X-ray gas is of the order of
  10 keV. This is very high when we are concerned with producing 3.5 keV
  photons, and we could say it is above threshold.

  So we might expect that for each collision there is an of the order
  unity chance of getting a large part of the energy of the hit made into
  electron  hole pairs, which in turn becomes hole electron annihilation
  photons, which are the 3.5 keV line photons to be observed
  may be on earth or better on a satellite near the earth.

  The intensity which we extracted from the data for this outskirts of the
  Perseus Cluster was
  \begin{eqnarray}
    \hbox{Intensity}_{PCO}&=& 5*10^{4
    1} ph*cm^6/s/SNe/GeV.
      \end{eqnarray}
\item{{\bf TSNR}} Thyco  Super Nova Remnant:

  The temperature in supernova remnants - at the time we now see
  Thyco Brahe's Supernova - is rather only 1 keV, so it already does not
  immediately guarantee that there is sufficient energy for production of
  3.5 keV photons. Rather some sort of good luck is needed, and we would expect
  that the rate could be appreciably lower.

  The rate is
  \begin{eqnarray}
    \hbox{Intensity}_{TSNR}&=& 1.4*10^{41} ph*cm^6/s/SNe/GeV
     \end{eqnarray}
  which is about 3 times lower.

  Since the $kT=1keV$ is lower than the $kT=10keV$ for the PCO, we could
  have expected an appreciable lower intensity
  $\hbox{Intensity}_{TSNR}$ than the $\hbox{Intensity}_{PCO}$
  due to some Boltzmann factor, but it seems that the two agree extremely well, since of course a factor 3 is far below our estimation accuracy.
  

\item{{\bf MWC}} Milky Way Center:

  Much gas in the Milky Way Center has very low temperatures like lower than
  250 K, although there also is diffuse plasma with $10^6$ to $10^7$ K.
  The collision velocity with dark matter will then order of magnitudewise
  be governed by the velocity of the dark matter which being of the order
  of 300km/s for hydrogen corresponds to a temperature 1 keV (about $10^7$ K).

  This means that the rate is expected to be even lower compared to the
  ideal one - say the one for PCO - than the one for TSNR.

  Indeed we have estimated the rate found for the Milky Way Center observed
  to be
  \begin{eqnarray}
    \hbox{Intensity}_{MWC}&=& 7.3*10^{38}ph*cm^6/s/SNe/GeV,
  \end{eqnarray}
  which is about 200 times smaller than for TSNR. If it - as it seems to -
  happened that it is in this step -from TSNR to MWC that the threshold for
  hydrogen and helium  collisions being able to produce 3.5 keV, then
  by this passage a fall in the 3.5 keV production should be by a factor
  equal to the ratio of hydrogen and helium abundance compared to the metals.
  This is of the order of 100.
  \item{{\bf DAMA}}

    We shall see in another section that the modulation part of the
    signal DAMA-LIBRA sees comes from the very tail of the tracks of the
    dark matter particles, just before they stop totally. This means that
    the effective velocity of the particles seen as modulation is very low.
    If we say that it is the remaining part of the track which is there only
    in one season, then since the velocity of the Earth around the sun is
    about 30 km/s, the velocity of this remaining track particles will be of
    the order of 30 km/s. At so low velocity the effective temperature in
    collisions will be for hydrogen 10 eV, and even for iodine with atomic
    weight $A_{iodine}=126.905\; amu$ only 1.27 keV. So even for heaviest
    atoms the velocity is below the threshold and only a very low
    fraction of the collisions are expected to produce 3.5 keV radiation.

    But that is o.k. because we just claimed that of the kinetic energy
    for incoming dark matter only a fraction of one in half a milliard
    is turned into 3.5 keV radiation.
%
However this way of explaining away more observations in DAMA has the problem, that it suggests that there are with higher velocities more events that do not
vary seasonally. There are, in fact, limits on how much non-modulating
background there can be, because it should have been seen as electron recoil
background in experiments like LUX-Zeppelin.

LUX Zeppelin claimed that the bit of background spilled over from the
electron recoil into the WIMP region was 3.6 mDRU with an expectation
2.6 $\pm$ mDRU. Here a mDRU = $10^{-3}cnts/keV_{ee}/kg/day$. Now we can see on
the plot count that there are 20 spilled over events out of a total of 160 background events observed by LUX Zeppelin.
This means that the full background essentially of electron recoil events is $\frac{160}{20}*3.6mDRU =28.8 mDRU$.

\end{itemize}

\section{Depth and Location Dependence of Underground Dark Matter Signal}
\label{ast}
We have earlier called attention to the fact that dark matter having significant
interaction with the earth, through which it penetrates down to dark matter underground experiments, can cause a significant dependence of the signal on the depth of the experiment. This could potentially explain the fact that DAMA-LIBRA sees a signal while ANAIS and LUX Zeppelin do not see
the corresponding signal, which is an impossible
situation for simple WIMP models.

Assuming, as we have in the present article, that the main velocity dependence
of the ratio $\frac{``\sigma''}{m}$ is as the inverse square of the velocity and that a major fraction of the stopping energy when the stopping is inelastic
goes into the production of 3.5 keV radiation, we can argue in the following way:

We take that the quotation marks around the $\sigma$ means that a cylinder
of the medium being penetrated of cross section $``\sigma''$ is brought into a
velocity of the order of the velocity of the pearl inducing this motion.
This means that per unit distance penetration the momentum loss
of the pearl or the object considered is $``\sigma'' *\rho*v$. But now since
we take the ansatz
\begin{eqnarray}
	\frac{``\sigma''}{m} &=& K/v^2,
\end{eqnarray}
the loss of momentum comes to behave like
\begin{eqnarray}
	\hbox{`` p-loss per distance  unit''} &=&\rho*m *K/v\\
	\hbox{`` E-loss per distance  unit''}&=&v/2 * m\rho*K/v= \rho*m*K/2\\
	\hbox{i.e.} ´´force''&=&\rho*K*m/2\\
	\hbox{and }``acceleration'' &=& \rho*K/2
\end{eqnarray}
For such a constant force the energy deposited per unit length becomes the same
all along the stopping path, and if the efficiency for making it into
the 3.5 keV radiation is constant too, then the stopping track will radiate
equally much per length unit whatever the speed of the particle causing
this radiation. So whether a dark matter particle enters with low or high
velocity will only become visible for instruments observing the 3.5 keV
radiation, when the particles stops. Then namely a fast particle still goes
on a little longer than the slow one. So in the here sketched approximation
the seasonal effect modulation can be calculated as if only the end tip of the
stopping track counts.

In the approximation that the motion of the Earth around the sun is small
compared to the motion of the solar system in the Milky Way we would just
get infinitesimally small pieces of track to count at the end of the track, i.e.
where the particles stop. The density of such stopping points in the Earth
would just reflect the kinetic energy distribution of the vertical direction,
since with our assumption of the force on the particle during the stopping
being constant, the length of the track in the vertical direction would
just be proportional to the kinetic energy along this direction.
(Actually we expect that thinking of a splitting up of the kinetic
energy according to coordinate axes is at least crudely o.k.)

\subsection{Motion of us  relative to Dark Matter}

The distribution of the dark matter in the galaxy is expected not to follow
the rotation of the visible matter, but rather being as a whole at rest
in the galaxy and only having a random Maxwellian distribution in velocity
\begin{eqnarray}
	f(\vec{v}) d^3\vec{v}&=&\left [\frac{m}{2\pi }\right ]^{3/2}
	\exp(-\frac{mv^2}{2T})d^3\vec{v}
\end{eqnarray}
where the temperature $T$ (with Boltzmann's constant $k=1$) should be adjusted
so as to make the dominant speed of the dark matter particles be around
250 to 280 km/s. I.e. that we should divide the total square of the
velocity of the order of $7.02*10^4km^2/s^2= 7.02 *10^{10}m^2/s^2$
into three equal portions to the three spatial coordinates. Thus we
shall have
\begin{eqnarray}
	\frac{2T}{m}&=& \frac{1}{3}*7.02*10^{10}m^2/s^2\\
	&=& 2.34*10^{10}m^2/s^2\\
	\Rightarrow \hbox{ ``typical velocity component '' }&=&
	\sqrt{ 2.34*10^{10}m^2/s^2}\\
	&=& 153 km/s = 1.53*10^5m/s
\end{eqnarray}

The direction of motion of the solar system
relative to the Milky Way is close to the direction of Deneb (=$\alpha$
Cygni), which has right ascension $20^h 41^m25.9^s$ and declination
$+45^016`49``$. Hence the velocity of 232 km/s in this direction will
have a component in the direction of the earth rotation axis oriented north
\begin{eqnarray}
	\hbox{``component north''}&=& 232km/s *\cos( 90^0 - ( +45^016`49``))\\
	&=& 0.7105*232km/s\\
	&=& 165 km/s= 1.65*10^5 m/s. 
\end{eqnarray}

\subsubsection{Laboratory on North Pole}

As an example, we consider first an underground laboratory placed on the North
Pole:

The distribution of the velocity component vertically down of the dark matter $D_{N.P.}(v_{vertical})$
would be a displaced Gaussian, though of course one cannot observe
up-going dark matter particles in a model wherein the dark matter gets
stopped on the km-scale:
\begin{eqnarray}
D_{N.P.}(v_{vertical})dv_{vertical}
	&\propto&
	\exp(-\frac{(v_{vertical}- 165\,km/s)^2}{(153\,km/s)^2})dv_{vertical}\nonumber\\
	&& \hbox{for } v_{vertical}
	\in [0\,km/s, \infty]
\end{eqnarray}
One should have in mind that the stopping depth will be rather simply
related to this incoming downward speed $v_{vertical}$. Also one should
have in mind that the speed of the Earth relative to the Milky Way varies
slightly with season, so that what we have put in as the velocity $165\, km/s$
varies with season with a fraction of the earth velocity around the sun,
which is 30 km/s.

For a detector at the depth corresponding to the stopping point for a vertical velocity
smaller than the peak in the distribution, here at the $165\, km/s$,
i.e. above where the particles of just this velocity stop, it should
see a seasonal variation opposite to that observed by the DAMA-LIBRA experiment.
However, for deeper experiments on the northern hemisphere one should get
the sign of the seasonal effect which DAMA-LIBRA got. This DAMA-LIBRA experiment
actually found that in the season when the Earth moved towards the bulk of the
(supposed) dark matter their event rate was higher than when the Earth moved
along the dark matter stream, escaping.

\subsubsection{Generalized Picture}

A similar consideration for the South Pole would just change the sign of the velocity
$165$ km/s and that would lead to that the deeper one goes down with one's
experiment, the seasonal effect will remain of opposite sign to that
found by DAMA-LIBRA. It would actually correspond to a derivative of the
tail of a Gauss distribution.

When one goes to the lower latitudes there should in principle be a variation
with the time of the day, but if one just averages that out by not measuring or
noting down the time on the day, then the approximate picture for the distribution
of the $v_{vertical}$ would be broader as one approaches the equator and the peak
velocity will be smaller also. At the equator there would be no seasonal effect
at the surface of the earth, the effect just at the surface changes sign when
passing the equator.

\includegraphics[scale=1.8]
{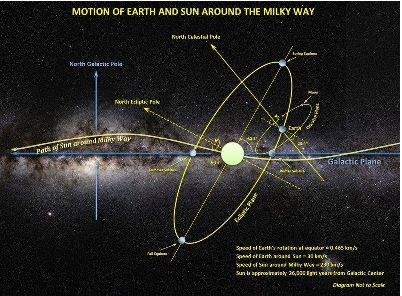}

2\includegraphics[scale=0.3]{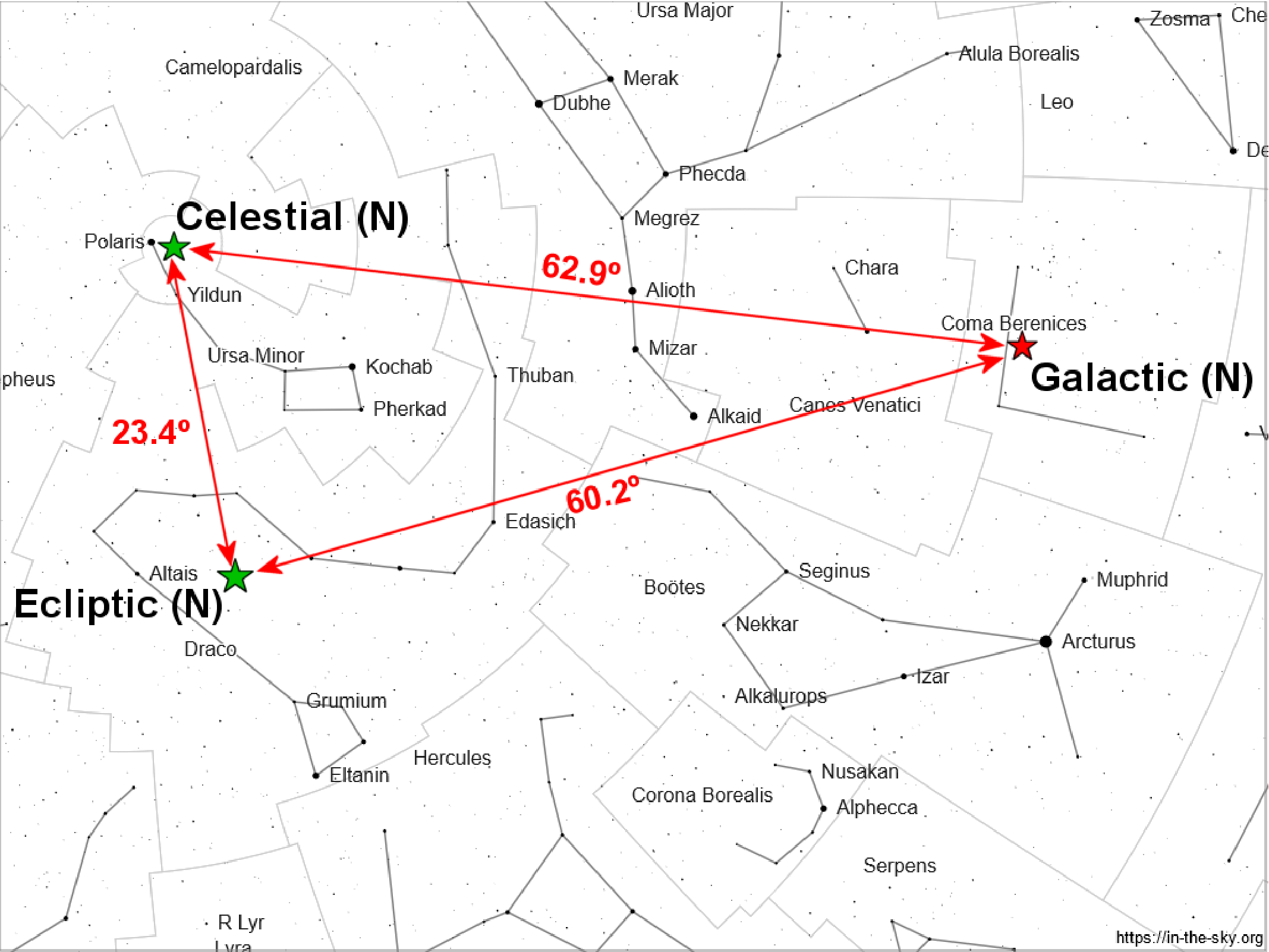}

\includegraphics[scale=2.1]
{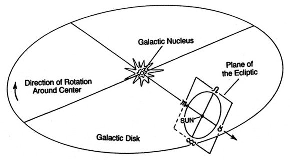}

\section{Conclusion}
We have re-looked at our since long announced model for
dark matter as being pearls of essentially ordinary matter
under very high pressure with correspondingly very
strongly screened nuclei in a dense degenerate electron
fermi gas, probably surrounded by a vacuum phase separating the
surface as suggested by Columbia plots.

Our main new point has been to use how the various nuclei
inside such a dark matter pearl interact individuality
or strongly interfering respectively in inelastic X-ray
producing collisions and elastic events. A major success
is that we can treat scattering of dark matter with dark
matter and with ordinary matter very similarly. In our
crude picture we solve or deliver chance of solving some
mysteries about the non-gravitational interactions of
dark matter:

\begin{itemize}
	\item The difficulty of fitting the 3.5 keV observations
	from the Perseus Cluster we propose to solve by
	allowing, that the outskirts of this galaxy cluster
	has its 3.5 keV radiation emitted from collisions of
	ordinary matter with dark matter processes.
	\item The mystery that only DAMA-LIBRA so far has seen the
	dark matter in underground direct detection, while
	others LUX have upper limits which in WIMP models
	seem quite contradictory, we solve by:
	
	The dark matter gets stopped in the earth with a stopping
	length of the order of the depth of DAMA, and what DAMA
	sees is really 3.5 keV radiation from dark matter
	having been excited by its passage through the earth
	above the experimental hall or in the apparatus. 
	
\end{itemize}

\subsection{Main Coincidence}

The main coincidence observed in the present article is that
the essentially stopping power per (weight per area) quantities
$\frac{´\sigma''}{m}$ are of similar order of magnitude for three
different processes, provided the velocity is higher than where threshold effects
would be expected to suppress the 3.5 keV radiation production. These three
different processes are:
\begin{itemize}
	\item The dark matter on dark matter scattering producing 3.5 keV X-rays.
	\item the dark matter on ordinary matter producing 3.5 keV X-rays.
	\item and the stopping of dark matter pearls on ordinary matter in the earth
	assumed to stop them around the depth of the DAMA experiment.
\end{itemize}

Then we come up with a story of interference that could allow the elastic
scattering of dark matter on dark matter being allowed to have much higher
cross section per mass than the three processes just mentioned. The Correa
interaction of dark matter with dark matter is measured in this way to be about half a million
times stronger than the three almost equal values of $\frac{\sigma}{m}$ mentioned above. This number of order $5*10^5$ is interpreted as the number of screened nuclei constituents in a dark matter
pearl. The size of the dark matter pearls estimated this way fits reasonably
well with earlier estimates, as well as with the story about the different
energy scales due to a fluctuating  lattice by one of us (see the present
volume of the Bled Proceedings). 

\subsection{On the Impact on the Earth}

We have given a series of predictions for how the chances of finding dark matter
in the underground by direct detection should vary between the hemispheres
of the Earth and with the depth.

\section{Appendix Translation of Units}

We here list the translations between the two different sets of units,
which we use, for the quantities of interest in this article:

Using
\begin{eqnarray}
	1 m &=& 5.07 *10^{15}GeV^{-1}\\
	1 kg &=& 5.625 *10^{26}GeV
\end{eqnarray}
we can write:
\begin{eqnarray}
	m^2/kg &=&4.57*10^4 GeV^{-3}\nonumber\\
	m^4/s^2/kg&=&  5.08*10^{-13}GeV^{-3}\nonumber\\
	kg/m^3 &=& 4.32*10^{-21}GeV^4\nonumber\\
	\hbox{earth density } 3000kg/m^3 &=& 1.29 *10^{-17}GeV^4\nonumber\\
	\hbox{DAMA depth } 1400 m &=&7.10*10^{18}GeV^{-1}\nonumber\\
	\hbox{DAMA depth $\times$ earth density}\, 4.2*10^6kg/m^2 &=& 91.6\, GeV^3
	\nonumber\\
	\hbox{inverse of this}\, 2.38 *10^{-7}m^2/kg &=&1.09*10^{-2}GeV^{-3}\nonumber\\
	\hbox{Typical gal. vel.}\, 3*10^5m/s&=& 10^{-3}
\end{eqnarray}

\end{document}